# Reconnection in a Pinch

T. E. Moore[1,3], J. L. Burch[2], D. E. Wendel[3]
1. 3rd Rock Research, Annapolis MD 21401
2. Southwest Research Institute, San Antonio, TX 78238
3. NASA Goddard SFC, Greenbelt MD 20771

\* Correspondence: T. E. Moore, thomem@iiirdrock.net

## Abstract

A recently published, new analysis of current sheets updated the classic Harris 1D static solution by considering multiple classes of charged particle trajectories in a generalized dynamic current sheet. It used a 1D PIC simulation to describe dynamic pinching and bifurcation of the sheet. These 1D results strongly suggest that plasma beta or other properties of the inflowing plasma have a strong effect on the equilibrium thickness of the pinched current sheet, but cannot describe magnetic reconnection. The time appears right to carry such 1D studies over to 3D simulations where current sheet thickness has been found exert an enabling or disabling influence on reconnection. The Magnetospheric Multiscale Mission (MMS), with its well-resolved multipoint measurements, has found that reconnection is enabled in collisionless plasma by non-adiabatic motions of electrons that can only occur in narrow magnetic structures with a scale comparable to electron inertial lengths ($d_e$). The recent 1D studies strongly suggest that a pinch to such scales can only occur for inflowing magnetized plasmas with relatively low plasma beta. We conclude that a parametric exploration of 3D simulated and observed inflow conditions, especially plasma beta, should shed light on the enablement of reconnection in collisionless plasmas.

## Introduction
Magnetic reconnection is the focus of the Magnetospheric Multiscale Mission (MMS), launched in March 2015 to investigate a process that had been invoked and identified for decades as an important factor in the storage and release of energy in the solar atmosphere and the magnetospheres of planets[1]. The design of the mission was driven by a belief, based on theoretical efforts, that phenomena on the smallest scales (electron gyro radius and inertial length or skin depth) are crucial to a full understanding the process.

The capabilities of MMS have revealed a cornucopia of small scale phenomena that provide a comprehensive view of the reconnection process in the all-important electron diffusion region where a strong electric field produces the plasma and magnetic flux transport of reconnection[2]. The reconnection process was revealed to be mostly laminar on the smallest scales, in accord with the generalized Ohm's Law (GOL), without dominant effects corresponding to dissipative turbulent fluctuations[3].

This is not to say that fluctuations and turbulence never play important roles in reconnection. However, it does seem to be that the most crucial factor is the tight curvature of the magnetic field in thin current sheets, leading to non-adiabatic motions that enable dissipation without collisions or noise[4,5]. The main point of this paper is that the factors that control current sheet thickness appear to be of great importance.

## Harris Current Sheet Thickness

The classical kinetic description of steady current sheets[6] yields a steady sheet when pressure balance is reached, that is when the peak plasma pressure is equal to the external magnetic field pressure ("peak plasma beta" of unity). The thickness of the Harris sheet ranges up to many times the plasma Debye length (MHD scales). However, when the magnetization electron drift speed of the pressure gradient is comparable to electron thermal speeds, the Harris current sheet thickness approaches the electron inertial scale or skin depth ($d_e = c/\omega_{pi}$). The magnetization drift speed increases with the steepness of the pressure gradient and the magnitude of the magnetic field, while the electron thermal speed is a free or at least an adjustable parameter. This suggests that low exterior plasma beta and thermal speed are consistent with a kinetic scale current sheet.

The simple Harris model of the statics of current sheets thus suggests that current sheets form in disequilibrium and then pinch or thin to an equilibrium thickness that may or may not be sufficiently thin to enable reconnection, depending on the prevailing plasma conditions on either side of the sheet current as it forms.

A relevant finding[7] argues that low beta current sheets collapse and increase the non-local "peak plasma beta" (ratio of plasma to external field pressure) as they do so. Similarly, Artemyev[8] found that current sheets in low beta plasmas collapse at a constant total current until $L_{cs} < d_e$, and that a dissipative reconnection electric field then appears.

## Yoon Current Sheet Thickness

Yoon et al.[9,10] studied the formation and development of current sheets under the action of the pinch effect, with a focus on their evolution into bifurcated diamagnetic reconnecting current sheets. Based on their analysis, they suggested that:

"… an initially thick, under-heated current sheet equilibrates to a thin, sub-skin depth bifurcated structure, which then undergoes collision-less reconnection … the equilibrium presented here [seems more likely] than widely-used specific solutions such as the Harris sheet."

Yoon et al. should be credited with having evolved and improved on the Harris current sheet model by considering the more complex non-adiabatic motions of particles in such current sheets, bringing to bear the work of Speiser[4,5]. Despite their focus on the bifurcation of thin current sheets, they have opened a path to exactly the studies that are needed to better understand where and when reconnection will be active or inactive in a given magnetized plasma situation. We briefly recap their results below:

They used the Harris solution as a guide, noting that, to be an equilibrium solution of the stationary Vlasov kinetic plasma, the drifting Maxwellian formulation must satisfy two important conditions:

1. The peak plasma pressure must equal the external magnetic field pressure (peak plasma beta = 1)
2. The sheet thickness must equal the Debye length times a factor of c/Vd, where Vd is the Maxwellian magnetization drift speed.

In general, there are multiple classes of particle orbits that take place in a current sheet, depending on its thickness, electric potential structure, and the presence of a guide field. These include non-crossing and crossing double-well potential orbits, partial crossing orbits and meandering orbits that are free of the potential influence. The relative population of the different orbit types is determined by the drift and thermal velocities of particles injected into the sheet, which may or may not satisfy the Harris equilibrium conditions above.

If a set of particles is inserted that is too cold or low beta to satisfy the Harris equilibrium ("under-heated" per Yoon et al. terminology), a linearized time dependent Vlasov analysis performed by Yoon et al. shows that the sheet particles are brought closer to the centre of the sheet and accelerated or heated. This was then simulated in full kinetic detail with consideration of all particle trajectory classes. The simulation started with a plasma temperature 0.2 of the Harris equilibrium temperature, and the initial thickness was $10d_i$. After initiation, the sheet was found to pinch and heat up for 30 ion cyclotron periods, after which it steadied at an equilibrium thickness of about $10d_e$. During this simulation, the particles were found to develop distinct non-Maxwellian features reflecting the evolution into different classes of orbit as the sheet pinched.

The final sheet was compared with observations from MMS of 17 June 2017:

"The half-width of the simulated sheet pinches down to ~$0.1\lambda = 1d_i = 10d_e$ (because $m_i/m_e = 100$ was used in the simulation).... The simulated sheet and the observed sheet are thus similar in that their widths are ~$10d_e$ and ≲ $1d_i$, so their sheet dynamics are mainly controlled by electrons."

These results were illustrated in Figure 4 of Yoon[9], which is reproduced here as Figure 1:

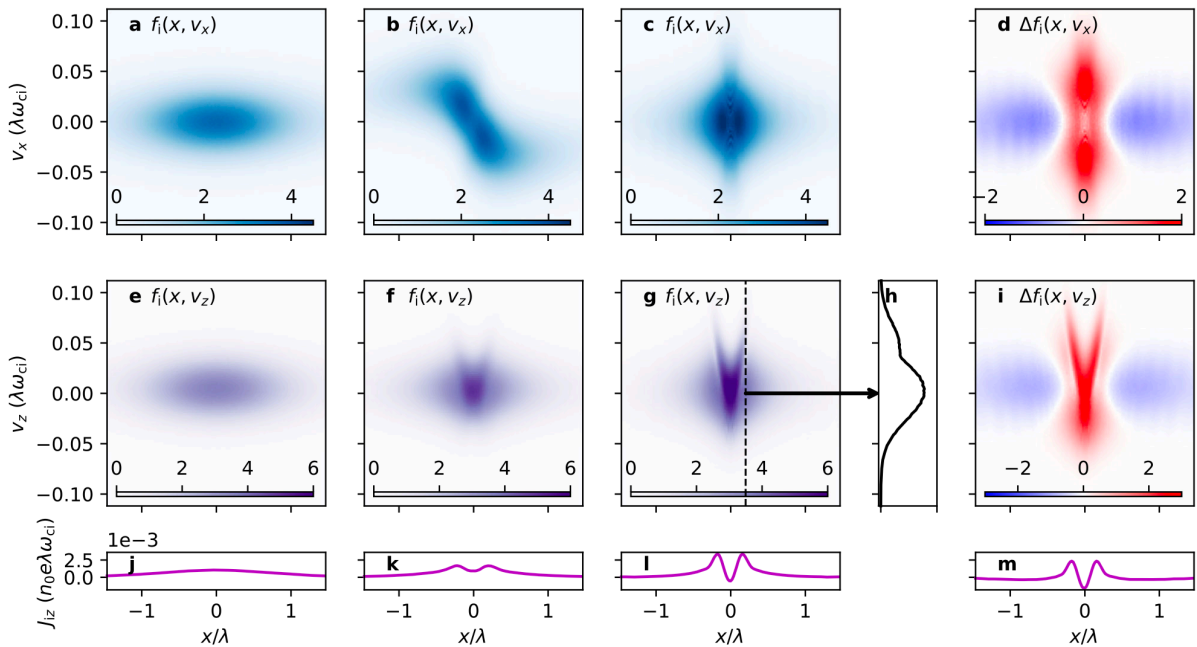

Figure 1. Evolution of particle velocity distributions and current density from an initial MHD scale with Maxwellian particles, to a bifurcated electron scale structure with non-Maxwellian features, in tens of ion cyclotron periods. Ion velocity distribution $f_i(x-v_x)$ and $f_i(x-v_z)$ space at **a**: $t=0$ $\omega_{pi}^{-1}$, **b**: $t=10$ $\omega_{pi}^{-1}$, and **c**: $t=100$ $\omega_{pi}^{-1}$. Difference (**d**) between $f_i$ at $t=0$ $\omega_{pi}^{-1}$ and $t=100$ $\omega_{pi}^{-1}$. **h** is a slice along dotted line in **g**. **j-l** are the current density $j_{iz}$, from **e-g**. **m** is the difference between **j** and **l**. [after Yoon et al.[9]].

Differences from the MMS observations were explained as the result of reconnection observed in the MMS case, whereas this could not happen in the simulated current sheet, since it was one dimensional. There was also substantial discussion of the simulated bifurcation of the pinched electron scale current sheet, which has also been observed in a number of cited direct space observations of both reconnecting and non-reconnecting current sheets. Overall, the simulations reflected closely the scenario quoted at the beginning this section, in which an initially thick current sheet thins to an equilibrium that was thin enough for reconnection to proceed, as observed by MMS.

As illustrated schematically in Figure 2., the Yoon et al.[9] results suggest that current sheets begin to form out of equilibrium at ~MHD scales and then undergo a pinch until they reach the equilibrium of field and plasma pressure balance. The final scale is determined by the amount of thinning required to raise the peak plasma beta to unity. The lower the initial plasma beta, the more thinning must occur for this to be achieved. For too high initial plasma beta, peak plasma beta of unity will be achieved before the sheet thins to electron scales.

As suggested by Figure 2, it seems reasonable to conclude that reconnection will not occur unless the inflowing plasma beta is sufficiently low to produce pinch thinning that approaches $d_e$.

Conversely, a reconnecting sheet in which the peak plasma beta were to be pushed higher than unity would subsequently thicken and disable reconnection. This could give rise to interesting dynamic variations of reconnection in response to inflow conditions or perhaps as a spontaneous instability even under steady external conditions.

It would be of great interest to explore the development of current sheets along the lines of the Yoon et al.[9] study, over a wide range of initial current sheet conditions, to determine quantitatively what parameter space of initial boundary conditions lead to thinning to kinetic scales. Also, it would be of even greater interest to explore how fully-developed, reconnecting current sheets react when the inflow conditions vary, using suitable multidimensional simulations. Presumably, conditions that would not support thinning and activation of reconnection would cause reconnection to be disabled by thickening of the current sheet to a new equilibrium. However, overshoots and oscillations around equilibrium were not seen by Yoon et al.

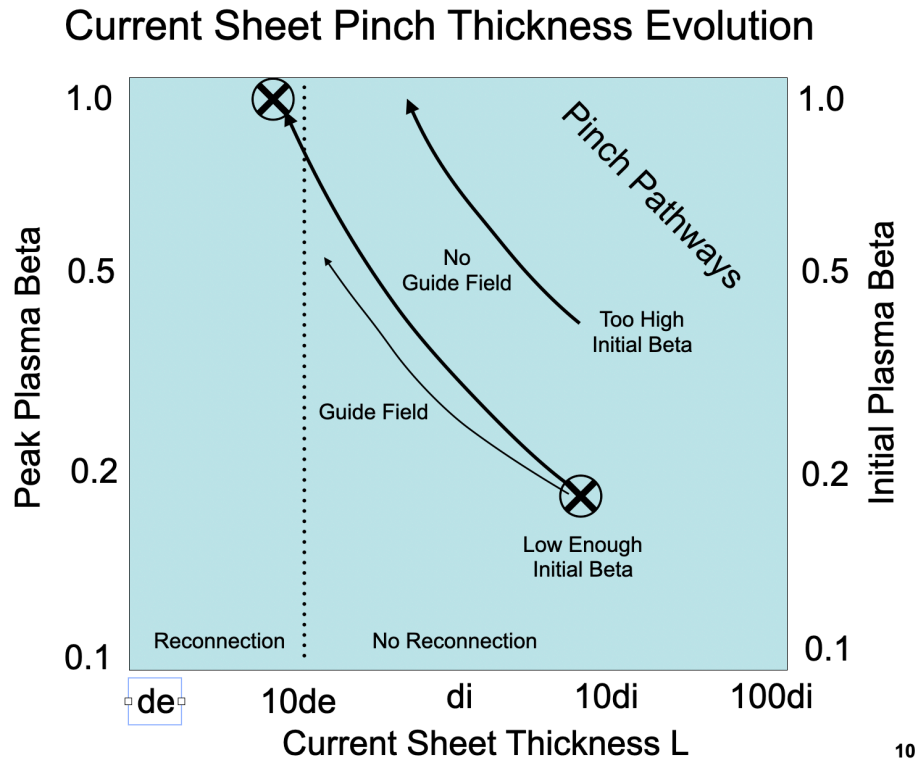

*Figure 2. Schematic extension of the current sheet study of Yoon et al.[9] illustrating pinch effect pathways during thinning and heating of MHD current sheets to kinetic scales, enabling reconnection for paths that reach thicknesses approaching $d_e$, and not for paths that terminate, that is, reach peak plasma beta of unity, at greater thickness. Points marked with circled X are those of the Yoon et al.[9] study.*

It is also tempting to speculate about magnetospheric circulation of higher and lower pressure plasma populations and how they might interact with reconnection, but that should perhaps be reserved for study using global circulation models with embedded, multidimensional kinetic descriptions that reflect the dynamics of current sheet thickness..

Yoon et al.[10] have just published new results that bear on the application to component or guide field cases. In this new paper, the role of a guide field is considered using the same tools as those described above. A guide field gives rise to the possibility of a mixed Harris and "force-free" equilibrium, in which the plasma pressure and peak plasma beta need not be raised as much by the pinch effect to reach a mixed equilibrium. That means less thinning of an initial MHD scale current sheet. Basically, less magnetic field shear means a smaller relevant reconnecting component of the magnetic field, which leads to a smaller amount of thinning. To force a guide field current sheet to reach kinetic scales and reconnect will apparently require lower initial beta plasma than for an antiparallel sheet, as illustrated in Figure 3. This should of course be explored in full multi-dimensional simulations of spontaneous reconnection.

## Future Needs

The activation or disablement of reconnection is known to be a fundamental control of energy release and transport in all of Heliophysics. We suggest that the knowledge gained from MMS, used in conjunction with multidimensional simulation tools[11,12,13,14,15], can be used to more fully explore current sheet formation and evolution and to elucidate the as yet unsolved problem of what initial boundary conditions produce spontaneous reconnection, leading to large scale system dynamics in reconnecting magnetized plasma situations.

We suggest a program containing some or all elements of the following:

1. Uniform unstructured boundary conditions of varying characteristics
2. Avoidance of features known to forcibly drive reconnection
3. Search for boundary conditions where reconnection is spontaneous
4. Search for boundary conditions that disable reconnection spontaneity
5. Separately investigate perturbations that initiate or disable reconnection

## Acknowledgements

The authors benefitted from enlightening discussions with Li-Jen Chen, Michael Hesse, Yi-shin Liu, Jan Egedahl, Mikhail Sitnov, and Masaaki Yamada. Figure 1 was obtained under open access to Nature Communications.

## Author Contributions

TEM conceived and developed the paper's theme and wrote the manuscript. JLB provided knowledge of cited papers as coauthor, reviewed and commented on the manuscript. DEW provided knowledge of cited papers as coauthor, reviewed and commented on the manuscript.

## References

1. Burch, J.L., Moore, T.E., Torbert, R.B. et al., 2016, Magnetospheric Multiscale Overview and Science Objectives. Space Sci Rev 199, 5–21. https://doi.org/10.1007/s11214-015-0164-9

2. Burch, J. L., and Phan, T. D., 2016, Magnetic reconnection at the dayside magnetopause: Advances with MMS, Geophys. Res. Lett., 43, 8327–8338, doi:10.1002/2016GL069787.

3. Torbert, R.B., T. D. Phan, M. Hesse, M. R. Argall, J. Shuster, R. E. Ergun, L. Alm, R. Nakamura, and Y. Saito, 2018, Electron-scale dynamics of the diffusion region during symmetric magnetic reconnection in space, SCIENCE, 15 Nov 2018, Vol 362, Issue 6421, pp. 1391-1395, DOI: 10.1126/science.aat2998

4. Speiser, T. W., 1965, Particle trajectories in model current sheets: 1. Analytical solutions, First published: 1 September 1965 https://doi.org/10.1029/JZ070i017p04219

5. Speiser, T.W., 1970, Conductivity without collisions or noise, Planetary and Space Science, Volume 18, Issue 4, April 1970, Pages 613-622

6. Harris, E., 1962, On a Plasma Sheath Separating Regions of Oppositely Directed Magnetic Fields, Nuovo Cimento 23, 115.

7. Takeshige, Satoshi, Shinsuke Takasao, and Kazunari Shibata, 2015, A Theoretical Model of a Thinning Current Sheet in the Low β Plasmas, The Astrophysical Journal, 807:159 (9pp), 2015 July 10 doi:10.1088/0004-637X/807/2/159

8. Artemyev, A.R., 2008, Evolution of a Harris Current Sheet in an Electric Field, ISSN 0027-1349, Moscow University Physics Bulletin, 2008, Vol. 63, No. 3, pp. 193–196. © Allerton Press, Inc., 2008. Original Russian Text © A.V. Artemyev, 2008, published in Vestnik Moskovskogo Universiteta. Fizika, 2008, No. 3, pp. 45–48.

9. Yoon, Young Dae, Gunsu S. Yun, Deirdre E. Wendel & James L. Burch, 2021, Collisionless relaxation of a disequilibrated current sheet and implications for bifurcated structures, Nat. Comm. | (2021)12:3774 |, https://doi.org/10.1038/s41467-021-24006-x

10. Yoon, Y.D., Wendel, D.E. & Yun, G.S., 2023, Equilibrium selection via current sheet relaxation and guide field amplification. *Nat Commun* **14**, 139 (2023). https://doi.org/10.1038/s41467-023-35821-9

11. Egedal, J., J. Ng, A. Le, W. Daughton, B. Wetherton, J. Dorelli, D. Gershman, and A. Rager, 2019, Pressure Tensor Elements Breaking the Frozen-In Law During Reconnection in Earth's Magnetotail, Phys. Rev. Lett. 123, 225101 – Published 25 November 2019

12. Hesse, M., Aunai, N., Birn, J. et al., 2016, Theory and Modeling for the Magnetospheric Multiscale Mission. *Space Sci Rev* **199**, 577–630, https://doi.org/10.1007/s11214-014-0078-y

13. Ji, H., Daughton, W., Jara-Almonte, J. et al., 2022, Magnetic reconnection in the era of exascale computing and multiscale experiments. Nat Rev Phys 4, 263–282 (2022). https://doi.org/10.1038/s42254-021-00419-x

14. Giovanni Lapenta1,2, Martin Goldman3, David L. Newman3, and Stefan Eriksson4, 2022, Formation and Reconnection of Electron Scale Current Layers in the Turbulent Outflows of a Primary Reconnection Site, The Astrophysical Journal, Volume 940, Number 2, DOI 10.3847/1538-4357/ac98bc

15. P. Sharma Pyakurel1, M. A. Shay1, T. D. Phan2, W. H. Matthaeus1, J. F. Drake3, J. M. TenBarge4, C. C. Haggerty5, K. G. Klein6, P. A. Cassak7, T. N. Parashar1, M. Swisdak3, and A. Chasapis1, 2019, Transition from ion-coupled to electron-only reconnection: Basic physics and implications for plasma turbulence, Physics of Plasmas 26, 082307 (2019); https://doi.org/10.1063/1.5090403